\documentclass[reprint,
usenatbib, 
amsmath,amssymb,aps]{revtex4-2}
\usepackage{subcaption}

\usepackage[dvipsnames]{xcolor}
\def\begeq{\begin{equation}}
\def\endeq{\end{equation}}
\def\begeqar{\begin{eqnarray}}
\def\endeqar{\end{eqnarray}}
\def\begeqal{\begin{align}}

\def\bfx{{\bf x}}
\def\bfr{{\bf r}}
\def\bfrhat{{\hat\bfr}}

\def\calY{{\cal P}}
\def\calZ{{\cal Z}}

	\newcommand{\calL}{{\cal L}}

	\newcommand{\six}[6]{\left(\begin{array}{ccc}
									{#1}& {#2}& {#3}\\
									{#4}& {#5}& {#6} \\
											  \end{array}\right)}

\begin{document}
\title{A Test for Cosmological Parity Violation Using the 3D Distribution of Galaxies}

\author{Robert N. Cahn}
\email{rncahn@lbl.gov}
\affiliation{Lawrence Berkeley National Laboratory, 1 Cyclotron Road, Berkeley, CA 94720, USA\\}

\author{Zachary Slepian}
\author{Jiamin Hou}
\affiliation{Department of Astronomy, University of Florida, 211 Bryant Space Science Center, Gainesville, FL 32611, USA\vspace{-2pt}\\}

\date{Accepted XXX. Received YYY; in original form ZZZ}

\date{\today}
\begin{abstract}
We show that the galaxy 4-Point Correlation Function (4PCF) can test for cosmological parity violation. 
The detection of cosmological parity violation would reflect previously unknown forces present at the earliest moments of the Universe. Recent developments both in rapidly evaluating galaxy $N$-Point Correlation Functions (NPCFs) and in determining the corresponding covariance matrices make the search for parity violation in the 4PCF possible in current and upcoming surveys such as those undertaken by Dark Energy Spectroscopic Instrument (DESI), the $Euclid$ satellite, and the Vera C. Rubin Observatory (VRO).
\end{abstract}
						  
\maketitle

\section{Introduction}

Among the known fundamental forces, only the weak interaction violates parity \citep{Wu:1957,Garwin:1957,Friedman:1957}. Since the weak interaction played no role in the evolution of the large-scale distribution of matter, observation of cosmological parity violation would imply the existence of new forces at the time of inflation. The Sakharov conditions \citep{Sakharov1967} for producing the baryon-antibaryon asymmetry (see {\textit e.g.} \citep{Riotto:1999yt} for a review) require violations of both charge conjugation invariance (C) and of CP,  the combination of C with parity (P). The weak interactions violate CP as well as parity \citep{Christenson:1964}, but this is well-described by the Standard Model of particle physics and 
  cannot account for the observed baryon-antibaryon asymmetry. Whatever new CP-violating force is responsible for the asymmetry may violate parity, as well.

Searches for parity violation have a long history. In 1848, Louis Pasteur directly observed a parity asymmetry. He found that artificially synthesized tartaric acid crystals could be separated into two distinct groups by their shapes. The crystals in one group were mirror images of those in the other group. However, tartaric acid produced organically in grapes yielded crystals of only one group. This occurred because organic molecules contain tetravalent carbon and when the carbon atom is attached to four different atoms the result is a tetrahedral shape that is distinguishable from its mirror image. That our hearts are on the left side of the body must owe its ultimate origin to the presence of just one form of each amino acid and the absence of its mirror image. Indeed, looking for dominance of organic molecules with a single chirality has been used in searches for extraterrestrial life~(see {\textit e.g.} \citep{Glavin2020} for a review).

In this work, we present a novel means of testing parity invariance in 3D large-scale structure, relying on the same principle as Pasteur's original separation: in general, in 3D a tetrahedron and its mirror image cannot be superimposed. 

The possibility of parity violation in large-scale structure is independent of homogeneity and isotropy: a large jar of crystals of one of the forms of tartaric acid would be homogeneous and isotropic to the extent of its volume.  
On the other hand, parity violation in 3D detected in the 4PCF would be evidence for primordial non-Gaussianity, since the 4PCF produced by a purely Gaussian random field would be simply products of 2PCFs and hence parity-conserving. In addition the technique described here can be used to search for parity violation in a straightforward way in 5-point and higher correlation functions using the algebraic structures described in \cite{Cahn202010}.

Tests of parity invariance in the cosmic microwave background (CMB) have been discussed for more than two decades ($e.g.$  \cite{1999PhRvL..83.1506L, 2009PhR...480....1A,2017JCAP...07..034B,Bartolo:2012sd,Bartolo:2014hwa,2016PhRvD..94h3503S,2011PhRvD..83b7301K}) and carried out in \cite{2020PhRvL.125v1301M}. Parity violation might be observable, as well, in primordial gravity waves\citep{Masui:2017fzw}. Our proposal opens the search to an entirely new class of experiments: 3D large-scale structure surveys.
 \section{Possible Sources of Cosmological Parity Violation}
 
 Two frequently considered potential sources of cosmological parity violation are represented by the Lagrangian densities
 \begin{eqnarray}
     \calL \propto \phi F^{\mu\nu}\tilde F_{\mu\nu}\label{eq:1}
     \end{eqnarray} 
and
\begin{eqnarray}
     \calL\propto \phi R^{\mu\nu\sigma\lambda}\tilde R_{\mu\nu\sigma\lambda}.\label{eq:2}
 \end{eqnarray}
 In Eq. (\ref{eq:1}), $F^{\mu\nu}$ is the field strength of an Abelian field like that associated with electromagnetism \cite{1999PhRvL..83.1506L, 1998PhRvL..81.3067C} and $\tilde{F}_{\mu\nu}=\epsilon_{\mu\nu\alpha\beta}F^{\alpha\beta}$ is its dual field, with $\epsilon_{\mu\nu\alpha\beta}$ denoting the Levi-Civita tensor. In Eq. (\ref{eq:2}), $R^{\mu\nu\sigma\lambda}$ is the Riemann tensor of General Relativity and $\tilde R_{\mu\nu\sigma\lambda}=\epsilon_{\mu\nu\alpha\beta}R^{\alpha\beta}_{\sigma\lambda}$ is its dual. In both cases $\phi$ is some scalar field of relevance in the early Universe, such as the inflaton or quintessence. Now, in both cases, there is no parity violation if $\phi$ is constant in space and time, since then the terms of Eqs. (\ref{eq:1}) and (\ref{eq:2}) are total derivatives.  {Such terms do not contribute to the equations of motion since integration by parts removes them from the action.}

In both cases, parity violation leads to a preferred helicity for fluctuations, respectively in the gauge field for Eq. (\ref{eq:1}) and in the metric for Eq. (\ref{eq:2}). This in turn induces parity-violation in the correlations between the curvature perturbations, and ultimately in the subsequent correlations between density fluctuations, which seed the formation of the galaxies we may observe in surveys of the late-time Universe.  
\section{Possibility of Parity Violation in the CMB}
Previous considerations of cosmological parity violation have primarily focused on CMB. CMB polarization can be decomposed into even- and odd-parity components for each angular momentum, $\ell$. It is possible to form parity-violating observables from  the product of temperature and the parity-odd polarization or from the product of the opposite-parity polarizations \cite{1999PhRvL..83.1506L}.

The temperature bispectrum of the CMB would evidence parity violation if a scalene triangle appeared projected onto the observed sphere more often than its mirror image. These two shapes could not be superimposed simply by sliding along the surface of the sphere. However, two mirror-image triangles can always be superposed in the full three-dimensional space. 
More generally, in a $D$-dimensional space, parity-odd basis functions must be a function of more than $D$ position vectors
\citep{Philcox:2021eeh}.

\section{Searching for Parity-violation with the 4PCF}
To search for parity violation we separate the parity-conserving and parity-violating components of the correlation function between fractional density fluctuations $\delta({\bf r}) \equiv \rho({\bf r})/\bar{\rho} - 1$ at locations ${\bf r}_i, i=0,1,2,3$, where $\rho(\bfr )$ is the density and $\bar{\rho}$ its average. By homogeneity one of the positions can be taken as the origin. Without loss of generality we thus set $\bfr_0=0$ so that the 4PCF is a function of three vectors; we denote it $\zeta(\bfr_1,\bfr_2,\bfr_3)$.
By isotropy $\zeta$ must be invariant under simultaneous rotation of ${\bf r}_1,{\bf r}_2$ and ${\bf r}_3 $. The 4PCF thus depends on three radial distances, $r_1,r_2,r_3$ and the collection of angles defining directions $\bfrhat_1,\bfrhat_2,\bfrhat_3$.

While our treatment ignores the anisotropy introduced observationally by redshift-space distortions (RSD; for a review, see \citep{Hamilton1998}), these do not produce spurious parity violation. One can straightforwardly show that RSD, when the 4PCF is averaged over all orientations with respect to the line of sight, do not produce any parity-breaking signal.

In previous work (\cite{Cahn202010}) two of us showed how to construct a complete set of isotropic basis functions of an arbitrary number of unit vectors from products of spherical harmonics. If the spherical harmonics are viewed as representing angular momenta, doing so is simply a matter of combining a number of angular momenta to form a quantity with zero total angular momentum. Here we confine our discussion to isotropic functions of three unit vectors, which is the case of interest for the 4PCF.

The isotropic functions can also be created in a Cartesian representation by forming a scalar quantity from a collection of unit vectors. For example, from $\bfr_1,\bfr_2,
\bfr_3$ one can form $\bfr_1\cdot\bfr_2,\  (\bfr_1\cdot\bfr_2)(\bfr_2\cdot\bfr_3),\  \bfr_1\cdot(\bfr_2\times\bfr_3)$, etc.  
This ``direct'' approach is convenient only for low values of the angular momenta, that is, for few factors of dot or cross products. It is clear that scalars formed from an odd number of unit vectors are parity-odd. 

Written explicitly in spherical harmonics, our isotropic functions are
\begin{multline}
\calY_{\ell_1\ell_2\ell_3}(\bfrhat_1,\bfrhat_2,\bfrhat_3)\\
=(-1)^{\ell_1+\ell_2+\ell_3} \sum_{m_1,m_2,m_3}
\six{\ell_1}{\ell_2}{\ell_3}{m_1}{m_2}{m_3}\\
\times Y_{\ell_1m_1}(\bfrhat_1)Y_{\ell_2m_2}(\bfrhat_2)Y_{\ell_3m_3}(\bfrhat_3)
\label{eqn:basis_funcs}
\end{multline}
where the matrix is a Wigner 3-$j$ symbol. To complete the specification of the basis functions we label the $r_i$ by the order $r_1\leq r_2\leq r_3$. Thus $(\ell_1,m_1) $ corresponds to the shortest of the $r_i$, and so on. The triangular inequalities, $|\ell_1-\ell_2|<\ell_3<\ell_1+\ell_2$ are enforced by the 3-$j$ symbol. The parity of the overall state is odd if the sum of the $\ell_i$ is odd.

The spherical harmonics form a complete basis for functions on the sphere and so the $\calY_{\ell_1 \ell_2 \ell_3}$ are a complete orthonormal basis for isotropic functions of $\bfrhat_1, \bfrhat_2, \bfrhat_3$. Using the orthogonality properties of the 3-$j$ symbols, we have
\begin{eqnarray}
&\int d\bfrhat_1 d\bfrhat_2 d\bfrhat_3\;
\calY_{\ell_1\ell_2\ell_3}(\bfrhat_1, \bfrhat_2, \bfrhat_3)\calY^*_{\ell'_1\ell'_2\ell'_3}(\bfrhat_1, \bfrhat_2, \bfrhat_3)\nonumber\\
&\quad=\delta^{\rm K}_{\ell_1\ell'_1}\delta^{\rm K}_{\ell_2\ell'_2}\delta^{\rm K}_{\ell_3\ell'_3},
\end{eqnarray}
where $d \bfrhat_i$ is the differential solid angle and $\delta^{\rm K}_{\ell\ell'}$ is the Kronecker delta, unity if its subscripts are equal and zero otherwise.

It follows from the properties of the spherical harmonics and the Wigner 3-$j$ symbols that
\begin{eqnarray}
     \calY_{\ell_1\ell_2\ell_3}(-\bfrhat_1,-\bfrhat_2,-\bfrhat_3)&=(-1)^{\ell_1+\ell_2+\ell_3}\calY_{\ell_1\ell_2\ell_3}(\bfrhat_1,\bfrhat_2,\bfrhat_3)\nonumber\\
     &=\calY^*_{\ell_1\ell_2\ell_3}(\bfrhat_1,\bfrhat_2,\bfrhat_3).
 \end{eqnarray}
 Consequently, $\calY_{\ell_1 \ell_2 \ell_3}$ is real if $\ell_1+\ell_2+\ell_3$ is even and is imaginary if the sum is odd. Moreover, the parity-odd components are those for which $\ell_1+\ell_2+\ell_3$ is odd. A variety of useful algebraic relations among the $\calY_{\ell_1\ell_2\ell_3}$ are given in \cite{Cahn202010}.
 
The 4PCF can be expanded as
\begin{eqnarray}
\zeta(\bfr_1,\bfr_2,\bfr_3)&=\sum_{\ell_1 \ell_2 \ell_3} \calZ_{\ell_1 \ell_2 \ell_3}(r_1,r_2,r_3)\\\nonumber\
&\times\calY_{\ell_1 \ell_2 \ell_3}(\bfrhat_1,\bfrhat_2,\bfrhat_3).
\label{eq:5}
\end{eqnarray}
It follows from the properties of the $\calY_{\ell_1 \ell_2 \ell_3}$ that $\calZ_{\ell_1 \ell_2 \ell_3}$ is real if $\ell_1+\ell_2+\ell_3$ is even and imaginary if the sum is odd.
The expansion coefficient $\calZ_{\ell_1 \ell_2 \ell_3}$ is obtained by averaging over the continuous position $\bfx$ out from which $\bfr_1,\bfr_2,\bfr_3$ are measured:
\begin{align}
\label{eqn:Zhat}
&\calZ_{\ell_1 \ell_2 \ell_3}(r_1,r_2,r_3)\\
&=\int \frac{d^3\bfx}{V}\int d\bfrhat_1 d\bfrhat_2 d\bfrhat_3\; \hat{\zeta}(\bfr_1,\bfr_2,\bfr_3;\bfx)\calY^*_{\ell_1 \ell_2 \ell_3}(\bfrhat_1,\bfrhat_2,\bfrhat_3),\nonumber
\end{align}
where $V$ is the volume over which ${\bf x}$ ranges. In the integrand, $\hat{\zeta}(\bfr_1,\bfr_2,\bfr_3;\bfx)$ is the estimate of the 4PCF obtained by sitting at a point $\bfx$, {\textit i.e.} it is $\delta(\bfx) \delta(\bfx + \bfr_1) \delta(\bfx + \bfr_2) \delta(\bfx + \bfr_3)$; we are projecting this estimate onto the basis of $\mathcal{P}_{\ell_1 \ell_2 \ell_3}$ and then, with $\int d^3{\bf x}/V$, averaging over all possible centers $\bfx$.

The problem of  measuring efficiently the large-scale 3PCF in the distribution of galaxies was solved by the technique developed in~\citep{Slepian201506,Slepian201510,Portillo2018}, with extensions to 4PCF and higher by \citep{Philcox:encore}. We briefly outline the approach here.

In practice we have in place of the continuous distribution of density fluctuations $\delta(\bfr)$ the collection of discrete galaxy locations. As a first step, we choose a galaxy at an absolute position $\bf{x}_i$. Next, we bin the relative distances of its neighbors into spherical shells which we denote by $r_j^{b}$. We then expand the angular dependence in each shell in spherical harmonics as:
\begin{align}
    \delta(\bfx_i,r^b_j,\bfrhat)=\sum_{\ell, m} a_{\ell m}(\bfx_i,r^b_j)Y_{\ell m}({\bfrhat})
    \label{eqn:expansion}
\end{align}
where
\begin{align}
    a_{\ell m}(\bfx_i,r^b_j)=\sum_\alpha Y^*_{\ell m}(\bfrhat_\alpha).
\end{align}
The summation is over galaxies $\alpha=1,2,\ldots $  in the radial bin $r^b_j$ surrounding the galaxy at $\bfx_i$. 

Using Eq. (\ref{eqn:expansion}) for $\delta(\bfx_i,r^b_j,\bfrhat)$ and forming the product indicated by $\hat{\zeta}$ (defined below Eq. \ref{eqn:Zhat}), we then project onto the basis of $P_{\ell_1 \ell_2 \ell_3}$ (see Eq. \ref{eqn:basis_funcs}) and average over $\bfx_i$ (the discrete analog of $\int d^3{\bf x}/V$ in Eq. \ref{eqn:Zhat}). The result is
\begin{multline}
    \calZ_{\ell_1\ell_2\ell_3}(r^b_1,r^b_2,r^b_3)\\
    ={(-1)^{\ell_1+\ell_2+\ell_3}}\sum_{m_1 m_2 m_3}
    \six{\ell_1}{\ell_2}{\ell_3}{m_1}{m_2}{m_3}\\
 \times {\cal A}_{\ell_1m_1,\ell_2 m_2,\ell_3 m_3}(r^b_1,r^b_2,r^b_3),   
\end{multline}
where we have defined
\begin{multline}
    {\cal A}_{\ell_1 m_1,\ell_2 m_2,\ell_3 m_3}(r^b_1,r^b_2,r^b_3)\\ 
    =\frac 1N\sum_{i=1}^Na_{\ell_1 m_1}(\bfx_i,r^b_1)a_{\ell_2 m_2}(\bfx_i,r^b_2)a_{\ell_3 m_3}(\bfx_i,r^b_3)
\end{multline}
and there are $N$ galaxies in the survey.

While the proposed search for cosmic parity violation is simple to describe, its implementation presents some challenges, which we now briefly outline. One concern is the computational expense of determining the 4PCF. The problem of measuring higher-point correlations on large scales efficiently was solved by the technique described above \citep{Slepian201506,Philcox:encore}.

The observables are the $\calZ_{\ell_1 \ell_2 \ell_3}(r^b_1,r^b_2,r^b_3)$. Each is specified by three integers constrained by the triangular inequalities, and by three radial bins. \textit{A priori} there is no preferred scale for searching parity violation. Searches where each tetrahedron side is of the order of a few Mpc to one or two hundred Mpc seem reasonable. With $\ell_{\rm max}=5$, a bin width of $10$ Mpc and a maximal side length $r_j^b$ of $200$ Mpc, there are thousands of different $\calZ_{\ell_1 \ell_2 \ell_3}(r^b_1,r^b_2,r^b_3)$.

More daunting is the challenge of determining the covariance matrix among so many observables. An analytic expression for the covariance matrix can be obtained by assuming a Gaussian random density field when evaluating the appropriate expectation value of eight density fluctuations \citep{hou2021analytic}. This offers a smooth, invertible template, which can then be calibrated using a reasonably modest number of mock catalogs. An analytical covariance template can be used to mitigate the sampling fluctuations that occur when the covariance is simply drawn from a number of mock catalogs~\cite{Scoccimarro200012,Joachimi2016} and due to these fluctuations may even fail to be positive semi-definite~\citep{VanDriel1978}.

Another challenge of a practical analysis is that computing NPCFs from spectroscopic surveys inevitably requires the use of randomly distributed particles to which the data are compared (\textit{e.g.} \cite{LS93,K2000,SS98}). The randoms are necessary to compensate for the survey footprint and for variations in the galaxy density in both the angular and radial coordinates. These latter can arise from inhomogeneities in the depth of the imaging survey used to target the spectroscopy as well as from the need to mask out the Galactic plane and bright stars.

Furthermore, NPCF computation from a spectroscopic survey requires a radial selection function, which gives the expected galaxy number density as a function of redshift in the absence of clustering. In practice, this selection function must be inferred from the observed spectroscopic data themselves, since the imaging survey used to target does not have radial information. In imposing this inferred selection function on the randoms which are then used to correct for the survey geometry, particular care must be taken to avoid spuriously introducing parity violation.  

Finally, it is important to handle carefully any possible systematics introduced by the mechanical method the survey uses to place an optical fiber on a galaxy, such as potential under-sampling of highly dense regions as can be caused by the limitations of coverage by robotic fiber positioners in DESI. Methods to correct these effects at the level of the 2PCF and its Fourier-space analog the power spectrum do exist \cite{Burden:2016cba,Pinol:2016opt,Bianchi:2018rhn} but need to be developed for the 4PCF.

Already there are significant data sets (Baryon Oscillation Spectroscopic Survey (BOSS), extended Baryon Oscillation Spectroscopic Survey (eBOSS)) that can be analyzed in the fashion described above. However, experiments now underway (Dark Energy Spectroscopic Instrument, DESI) and others that will soon take data (\textit{Euclid}, Vera Rubin Observatory, VRO) will provide much more extensive data and will enable more stringent tests of parity violation.  
 
A simple model-free test is to compare the magnitudes of the parity-violating terms (``signal') to the ``noise'' given by the Gaussian random field covariance. However, much more powerful tests will be possible of models that predict relations between the various parity-violating terms.

\section{Summary}
In this work we have proposed a new method to search for parity violation, the first using 3D large-scale structure (LSS). This method relies on the fact that in 3D, tetrahedra, represented by the 4PCF, are the lowest-order shapes that cannot be rotated into their mirror images, and so the lowest-order shapes that can probe parity violation. We have outlined how measurement of the 4PCF in a basis of isotropic  functions can be made computationally feasible, and discussed some of the challenges that will need to be carefully addressed to enable a robust analysis. The coming years will offer ideal datasets for 3D parity-violation constraints with the 4PCF. Overall, the detection of a parity-violating signal in LSS would illuminate early Universe physics and perhaps even reveal physical processes beyond the Standard Model.

\section*{Acknowledgments}
We acknowledge extensive contributions to this work by O.H.E. Philcox. The work of RNC was supported in part by the Director, Office of Science, Office of High Energy Physics, of the U.S. Department of Energy under contract No. DE-AC02-05CH11231. ZS gratefully acknowledges Lawrence Berkeley National Laboratory for hospitality during some of the period of this work.


\bibliography{ref} 

\label{lastpage}

\end{document}